# Interpenetrated biosurfactant-biopolymer orthogonal hydrogels: the biosurfactant's phase controls the hydrogel's mechanics

Chloé Seyrig,[a] Alexandre Poirier,[a] Javier Perez,[b] Thomas Bizien,[b] Niki Baccile[a,*]

[a] Sorbonne Université, Centre National de la Recherche Scientifique, Laboratoire de Chimie de la Matière Condensée de Paris , LCMCP, F-75005 Paris, France

[b] Synchrotron SOLEIL, L'Orme des Merisiers Saint-Aubin, BP 48 91192 Gif-sur-Yvette Cedex

\* Corresponding author:
Dr. Niki Baccile
E-mail address: niki.baccile@sorbonne-universite.fr
Phone: +33 1 44 27 56 77

**Abstract**

Controlling the viscoelastic properties of hydrogels is a challenge for many applications. Low molecular weight gelators (LMWG) like bile salts and glycolipids, and biopolymers like chitosan and alginate, are good candidates for developing fully biobased hybrid hydrogels that combine the advantages of both components. Biopolymers lead to enhanced mechanics while LMWG add functionality. In this work, hybrid hydrogels are composed of biopolymers (gelatin, chitosan, alginate) and microbial glycolipid bioamphiphiles, known as biosurfactants. Besides their biocompatibility and natural origin, bioamphiphiles can present chameleonic behavior, as pH and ions control their phase diagram in water around neutrality under strongly diluted conditions (< 5 wt%). The glycolipid used in this work behaves like a surfactant (micellar phase) at high pH or like a phospholipid (vesicle phase) at low pH. Moreover, at neutral-to-alkaline pH in the presence of calcium, it behaves like a gelator (fiber phase). The impact of each of these phases on the elastic properties of biopolymers is explored by means of oscillatory rheology, while the hybrid structure is studied by small angle X-ray scattering. The micellar and vesicular phase reduce the elastic properties of the hydrogels, while the fiber phase has the opposite effect, it enhances the hydrogel's strength by forming an interpenetrated biopolymer-LMWG network.

**Introduction**

Single network (SN) hydrogels consist of an entangled or cross-linked network, generally composed of a polymer or a biopolymer swollen by water. Polymeric hydrogels are of great interest for biomedical applications,[1] soft electronics and sensors,[2] water treatment,[3] drug delivery[4] and food applications[5] among others, but often suffer from a lack of mechanical strength. The main reasons are heterogeneous distributions of cross-linking sites, varying molecular weights between crosslinks and the lack of stress dissipation mechanisms to prevent crack propagation.[6,7] Increasing the polymer concentration and/or cross-linking density can significantly improve the mechanical properties of SN hydrogels, but present major drawbacks: firstly for manipulation purposes, and secondly for use in applications. For instance, in the biomedical field, dense gels can inhibit nutrient transport or adversely affect cell mechano-transduction and finally disturb cellular behavior.[1]

Biopolymers are of great use due to their inherent properties, especially their bioactivity, degradability and biocompatibility. However, they comprise certain drawbacks (wider distributions of molecular weights, undefined chemical compositions and possible immune response depending on their sourcing), that are not exhibited by petro-derived polymers.[1] These issues have been addressed by mechanical reinforcement through the hybridization of two (or more) interpenetrating networks, since the first report by the Gong group on the fabrication of interpenetrated polymer networks (IPN).[6,8,9] The advantages of IPN over SN hydrogels include mechanical reinforcement, the ability to respond to external stimuli and the tuning of cell-material interactions. Many studies on multicomponent systems have been carried out since, emphasizing that the final storage modulus depends on the combination of the gelators selected.[10] Hybrid gels can be composed of two polymers, but also of a polymer and other self-assembled phases expected to provide functionality, such as micelles,[11] fibers[12] and vesicles.[13] The latter, for instance, are useful for encapsulation and release purposes, as illustrated by the work of Dowling *et al.*, who efficiently encapsulated and controlled the release of calcein in NaOA vesicles loaded in gelatin hydrogels.

Combining two different fibrous networks has also been shown to be an efficient method to address most of the challenges of SN hydrogels.[14–16] Cornwell and Smith conceived a hybrid system composed of a low molecular weight gelator (LMWG) and a polymer gelator (PG). They used NMR and circular dichroism methods to demonstrate the stability of the polymer gel and the pH-responsivity of the LMWG,[17] the latter playing a key role to retain the overall integrity of the gel. Van Esch *et al* reported a nice example of orthogonal self-assembly of a surfactant and a gelator within a two-component system, benefiting from both the strength and stiffness of the hydrogelator's fibers and dynamism due to the presence of cylindrical surfactant micelles of cetyltrimethylammonium tosylate (CTAT). Stubenrauch *et al.* adopted another approach and worked on the gelification of non-gelling lyotropic phases using low-molecular weight gelators.[18–20] Despite a mutual influence, there is no deep modification of the structures of either the Lα phase or the gel network in the gelled Lα phase, demonstrating that the surfactant and the gelator molecules self-assemble according to an orthogonal process. In these examples, the authors agree on showing the important role played by the amphiphile. However, each system requires a specific amphiphilic molecule to ensure the formation of a given phase, and most often, if polymers can be replaced by biobased macromolecules (synthetic or natural biopolymers),[1] amphiphiles are commonly selected for their function and not for their sustainability. The latter parameter is however of utmost importance for possible applications in the cosmetic, biomedical and food science fields.

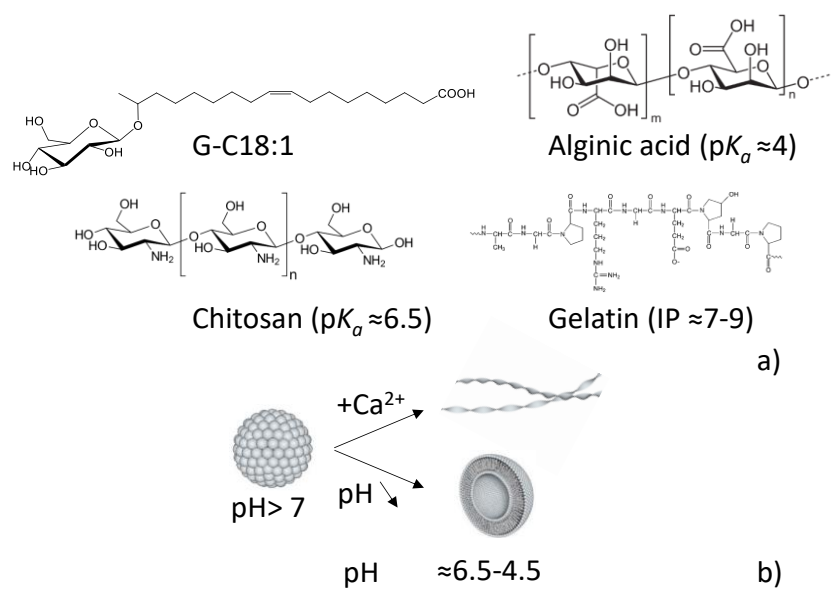

**Figure 1 - a) Chemical formulas of G-C18:1, alginic acid, chitosan and gelatin. b) Known phase behavior in water at room temperature (C< ~5 wt%) of G-C18:1. Micelles form at basic pH (counterions: $Na^+$) and undergo a micelle-to-vesicle transition when pH is lowered.[21,22] Crystalline fibers form from the micellar phase by adding $Ca^{2+}$.[23–25]**

In this work, we study the effect of various self-assembled phases, micellar, vesicular, fibrillar (Figure 1b) on the elastic properties of three biobased macromolecular hydrogels. All the phases are formed by one single biobased glycolipid amphiphile of microbial origin under close physicochemical conditions of pH and ionic strength.[21–24] The selected bioamphiphile, G-C18:1 (Figure 1a) is a fermented bolaform glycolipid composed of a single *β*-D-glucose hydrophilic headgroup and a C18:1 fatty acid tail.[26] Due to its double amphiphilic nature and its free-standing COOH group, G-C18:1 is negatively-charged under alkaline conditions. When negatively-charged, G-C18:1 forms a micellar phase[21,22] in the presence of $Na^+$ and a fiber phase with $Ca^{2+}$ (Figure 1b).[23,24] Reducing pH from alkaline to acidic (pH< ≈6.5) drives a micelle-to-vesicle phase transition (Figure 1b).[21,22] Recent work has shown its ability to enrich its phase diagram by strongly binding to cationic polyelectrolytes (e.g., poly-L-lysine or chitosan), but not to amphoteric polymers like gelatin under like-charge conditions, thus showing the complex multiphase behavior under diluted conditions in water of G-C18:1[27–29]

Considering the importance of the electrostatic interactions, the biopolymers selected for this work are gelatin,[30] chitosan and alginate (Figure 1a).[31,32] Not only do they have amphoteric, positive and negative properties, they are also among the most studied and commercially-exploited biopolymers.

Gelatin is obtained by the denaturation of collagen protein and has many potential applications due to its ability to stabilize colloids and its gelation below 30°C.[30] Below its isoelectric point, gelatin bears a positive net charge and is expected to interact with a negatively charged surfactant, resulting in the precipitation of polymer-surfactant complexes.[33–35] Alginate is a polysaccharide composed of $\beta$-D- mannuronic acid (M-blocks) and α-L- guluronic acid (G-blocks), as well as regions of interspersed M and G units.[31] Alginates form hydrogels in aqueous solution under mild conditions through interaction with divalent cations, mainly $Ca^{2+}$, whose cooperative binding between the G-blocks of adjacent alginate chains creates reversible ionic interchain bridges.[32] Finally, chitosan is obtained by the deacetylation of chitin, the second most widespread natural polysaccharide. Its structure involves both N-acetylglucosamine and glucosamine, linked together into linear chains through $\beta$-(1-4) connections.[36] Biocompatibility, biodegradability, anti-microbial activity and the promotion of wound healing are some examples of chitosan's advantages. Its structure is close to that of glycosaminoglycans, the main constituents of the natural extracellular matrix, thus opening up serious paths for developing tissue engineering applications.[37]

Using a combination of rheology and small angle X-ray scattering (SAXS), this work shows the enhancement of the elastic and loss moduli of biopolymer hydrogels upon mixing with a fibrillar phase of G-C18:1, and their loss when mixing with a micellar and vesicular phase.

**Experimental**

*Chemicals*

Glycolipid biosurfactant G-C18:1 was purchased from the Bio Base Europe Pilot Plant, Gent, Belgium, lot No. APS F06/F07, Inv96/98/99 and used as such. G-C18:1 ($M_w$ = 460 g.mol$^{-1}$) is monounsaturated and contains a β-D-glucose unit covalently linked to oleic acid. The molecule was obtained by fermentation from the yeast *Starmerella bombicola ΔugtB1* according to the protocol given before.[38] The molecular purity of G-C18:1 exceeded 95%. According to the specification sheet provided by the producer, the batch (99.4% dry matter) was composed of 99.5% of G-C18:1, according to HPLC-ELSD chromatography data. NMR analysis of the same compound (different batch) was performed elsewhere.[21] The three polymers used in this work, gelatin (type A, from porcine skin, $M_w$ ≈50–100 kDa, isoelectric point 7–9), alginate (from brown algae, medium viscosity, $M_w$ ≈20–240 kDa, p$K_a$≈4) and chitosan (high molecular weight, HMW, from shrimp shell, practical grade, $M_w$ ≈190–375 kDa, p$K_a$≈6.5), were purchased from Sigma Aldrich.

*Preparation of the hydrogels*

*Stock solutions.* The G-C18:1 stock solution at concentration, $C_{G-C18:1}$= 40 mg/mL, was prepared by dissolving the G-C18:1 powder in the appropriate volume of milli-Q water at pH= 8, adjusted with a few µL of NaOH 5 M or 1 M solution. The stock solution of G-C18:1 was clear and on the basis of previous work,[21,22] mainly micellar. After acidification (at pH< 6.5) of the stock solution by adding µL amount of 1 M HCl, the solution immediately became turbid. This indicated the transition towards a stable colloidal dispersion of vesicles in water. Further acidification below pH 4 led to the precipitation of a lamellar powder.[21,22]

The stock solutions for each biopolymer were prepared as follows. *Gelatin*: 80 mg of gelatin powder was dispersed in 2 mL of milli-Q water for a concentration of $C_{gelatin}$= 40 mg/mL. The gelatin stock solution was vortexed and set in the oven at 50°C. Once the solution was homogeneous, pH was increased to pH 8 with a few µL of a 0.5 M - 1 M NaOH solution. *Chitosan*: 200 mg of chitosan dispersed in 10 mL of 0.1 M acetic acid aqueous solution for a concentration C= 20 mg/mL. For optimal solubilization, the chitosan stock solution was stirred for one day before use. *Alginate*: 200 mg of alginate powder was dispersed in 10 mL of milli-Q water for a concentration C= 20 mg/mL and stirred until complete solubilization. The magnetic stirrer usually sticks upon water addition; in this case, manual assistance can be required to improve stirring. For a typical volume of 10 mL, pH is then increased to 10 with 1-10 µL of a 5 M or 1 M NaOH solution under stirring. Stirring and vortexing may be necessary to obtain a homogeneous alginate solution. Additional protocol details are given below for each biopolymer.

*Fibrillar G-C18:1 gels, {F}G-C18:1.* The preparation of fibrillar gels of G-C18:1 was described elsewhere.[23–25] Briefly, for a 1 mL sample, 500 µL of the G-C18:1 micellar stock solution was mixed with 500 µL of water. $CaCl_2$ solution (1 M, $V_{CaCl2}$= 33.5 µL, $[CaCl_2]$= 33.5 mM) was added manually for a total $[CaCl_2]$ : [G-C18:1]= 0.7 molar ratio. The final solution was stirred and a gel was obtained after resting a few hours at room temperature.

*{F}G-C18:1/gelatin gels.* A volume of 500 µL of the gelatin stock solution was mixed with 500 µL of either water (reference) or G-C18:1 stock solution (sample). For a typical volume of 1 mL, $CaCl_2$ solution (1 M, $V_{CaCl2}$= 33.5 µL, $[CaCl_2]$= 33.5mM) was added manually for a total $[CaCl_2]$: [G-C18:1]= 0.7 molar ratio. The final solution was stirred and a gel is obtained after resting a few hours at room temperature.

*{F}G-C18:1/chitosan gels.* The pH of 1 mL of the acidic chitosan stock solution was increased above about 8. Typically, for 1 mL, 5-10µL of a 5 M or 1 M NaOH solution was added. The initially viscous solution was vigorously stirred and vortexed to obtain a heterogeneous gel. Due to the heterogeneity of the chitosan gel at basic pH, the preparation protocol was slightly adjusted to avoid manipulating concentrated chitosan solutions. Instead of mixing volumes, 500 mg of the chitosan gel was weighed and mixed with 500 µL of a glycolipid solution under vigorous stirring and vortexing. 33.5 mM ($V_{CaCl2}$= 33.5 µL) of a $CaCl_2$ solution ($[CaCl_2]$= 1 M) was added to this mixture, followed by further mixing. The final concentration of $CaCl_2$ in the sample is 33.5 mM, for a total $[CaCl_2]$: [G-C18:1]= 0.7 molar ratio. A homogeneous gel was obtained after resting a few hours at room temperature. Note: a second method, consisting in employing $NH_3$ diffusion inside the acidic chitosan solution under a closed bell jar followed by water washing, was employed to prepare chitosan hydrogels. Although this approach provided more homogeneous gels, it was unsuitable for controlling the final pH and the diffusion of calcium. For these reasons, it was discarded.

*{F}G-C18:1/alginate gels.* 500 µL of the alginate viscous stock solution was added either to 500 µL water (reference) or to 500 µL of the G-C18:1 stock solution (sample) under stirring. A volume of $V_{CaCl2}$= 50 µL of a $CaCl_2$ solution ($[CaCl_2]$= 1 M) was added, for a final $CaCl_2$ concentration of 50 mM and $[CaCl_2]$ : [G-C18:1]= 0.9 molar ratio and $[CaCl_2]$ : [alginate]≈ 0.0015 molar ratio. The final solution was magnetically stirred for several hours to obtain a homogeneous gel.

Table 1 summarizes the reference, sample and stock solution concentrations in wt% employed throughout this study. Please note that the concentration of each biopolymer was adapted from system to system for practical use (chitosan and alginate, in particular, become highly viscous to the point of being hardly usable).

**Table 1 – Concentration of stock solutions, volumes from stock solutions and final concentration in the 1 mL samples.**

|  | G-C18:1 | gelatin | alginate | chitosan |
|---|---|---|---|---|
| $C_{Stock\ solution}$ / wt% | 4 | 4 | 2 | 2 |
| V / mL | 0.5 | 0.5 | 0.5 | 0.5 |
| $C_{Control}$ or $C_{Sample}$ / wt% | 2 | 2 | 1 | 1 |

*Rheology*

Viscoelastic measurements were carried out using an Anton Paar MCR 302 rheometer equipped with parallel titanium or stainless-steel sandblasted plates (diameter = 25 mm, gap = 1 mm). Unless otherwise stated, all the experiments were conducted at 25 °C, whereas the temperature was controlled by the stainless-steel lower Peltier plate. During the experiments, the measuring geometry was covered with a humidity chamber to minimize water evaporation. Dynamic oscillatory and time sweep experiments were performed by applying a constant oscillation frequency (f = 1 Hz) and a shear strain ($\gamma$ = 0.1 %) within the linear viscoelastic regime (LVER).

*Small angle X-ray scattering (SAXS)*

SAXS experiments were performed at 25 °C at the Swing beamline at the Soleil synchrotron facility (Saint-Aubin, France). The samples were analyzed during run No. 20201747 using a beam at 12.00 keV and a sample-to-detector distance of 2.00 m. The samples were prepared *ex situ*, pushed through a 1 mm quartz tube with a 1 mL syringe and analyzed directly by setting them in front of the X-ray beam. The signal of the same quartz tube containing water was subtracted as background. The quartz tubes were rinsed with water and ethanol after each use. The signal was integrated azimuthally to obtain the I(q) vs. q spectrum (q= $4\pi \sin \theta/\lambda$, where $2\theta$ is the scattering angle) after masking systematically wrong pixels and the beam stop shadow. Silver behenate ($d_{(001)}$ = 58.38 Å) was used as SAXS the standard to calibrate the q-scale. Data were not scaled to absolute intensity.

*Rheo-SAXS*

Experiments coupling rheology and SAXS were performed at the SWING beamline of the Soleil synchrotron facility (Saint-Aubin, France) during run No. 20200532, using a beam energy of 12.00 keV and a sample-to-detector distance of 1.65 m. Tetradecanol ($d_{(001)}$ = 39.77 Å) was used as the q-calibration standard. The signal was integrated azimuthally with Foxtrot software to obtain the I(q) spectrum (q =4π sin θ/λ, where 2θ is the scattering angle) after systematically masking defective pixels and the beam stop shadow. A MCR 501 rheometer (Anton Paar, Graz, Austria) equipped with a Couette polycarbonate cell (gap= 0.5 mm, volume, ≈ 2 mL) was coupled to the beamline and controlled through an external computer in the experimental hutch using the Rheoplus/32 software, version 3.62. The experiments were performed in a radial configuration, where the X-ray beam is aligned along the center of the Couette cell. The data were not scaled to absolute intensity.

*Measuring critical micelle concentration (CMC) by surface tension*

CMC was determined by plotting the static surface tension of a G-C18:1 solution against the G-C18:1 concentration. The surface tension was measured with the pending drop method by means of a drop shape analysis system DSA30 Krüss, Germany, with associated software and 1 mL borosilicate glass SY20 microsyringes. A pendant drop of 11 – 30 mL of the solution was produced in air with a steel capillary having an external diameter of 1.83 mm. Images were recorded every 0.4 s for 20 s and averaged (50 values) to obtain a value of the surface tension. The contour of the drop was fitted by the Young-Laplace equation using an iterative process with the surface tension, γ, as an adjustable parameter. The cleanliness of the setup was verified by pumping the syringe volume with milliQ water 10 times. The surface tension had to be constant and reproducible ± 0.5 mN/m during the total time of the experiment. The CMC was measured at pH 10 and pH 6.5, when G-C18:1 was in the micellar and vesicle phase, respectively. Experimentally, the G-C18:1 solution at the highest concentration was diluted with a water solution at the same pH, 10 or 6.5.

**RESULTS**

G-C18:1 has a rich phase behavior in water at room temperature.[29] Its pH-dependent self-assembly, summarized in Figure 1b,[22] shows that micelles (diameter, d ≈5 nm) and unilamellar vesicles (membrane thickness, T ≈5 nm and polydisperse outer diameter above 50 nm) form above and below pH ≈6.5, respectively,[21,22] at the same concentration and temperature. Neither one of them has any gelling property in the concentration range employed in this work. An additional phase has recently been revealed when adding metal ions and in particular a $Ca^{2+}$ salt to the micellar phase (pH above ≈7.8), keeping both temperature and concentration strictly equal. Self-assembled, infinitely long fibers with a cross-section of about 10 nm spontaneously and immediately form, as shown by cryo-TEM and their characteristic SAXS profile.[23–25] Fibrillation occurred homogeneously, followed by prompt gelling. The typical storage modulus, G', of G-C18:1 hydrogels settled around 100 Pa at about 2 wt% for an equimolar amount of positive-to-negative charge, corresponding to a molar ratio of $[CaCl_2]/[G$-C18:1$] > ≈0.5$ (Figure S 1a).[25]

The viscoelastic behavior of all the biopolymers studied in this work is given in Figure S 2 for selected controls at two pH values in the form of storage and loss moduli measured a few minutes after loading the rheometer. As expected, all the reference samples were gels of which the magnitude of the elastic modulus depended on the nature of the biopolymer, but not on the pH.

*Hybrid {F}G-C18:1/gelatin hydrogels*

Network interpenetration between polymer gels and self-assembled fibrillar network (SAFiN) hydrogels were able to generate a synergistic interaction, improving the elastic performance of the gels. This was studied for gelatin {F}G-C18:1 SAFiN hybrid hydrogels in Figure 2 and Figure 3. As can be seen in Figure 2, the aqueous vesicular and micellar solutions mixed with a 2 wt% gelatin have comparable G' (~ 30 Pa) and do not have significant differences in the final properties of the gel, although they seem to have a moderately detrimental impact on G', if compared to gelatin alone (~ 55 Pa, Figure S 2). On the other hand, the addition of a $Ca^{2+}$ solution to the gelatin-micelle gel strongly improves the corresponding G' by almost one order of magnitude. The repeatability of the experiment shows about a 4% error across multiple freshly-prepared samples (G' = 205 ± 9 Pa averaged over 5 samples, Figure S 3). The mechanical properties of the hybrid gelatin-SAFiN gel were also better than those of each component alone, {F}G-C18:1 and gelatin, compared in Figure 3a, thus showing how the combination of both fibrillar and polymeric components generates a stronger gel.

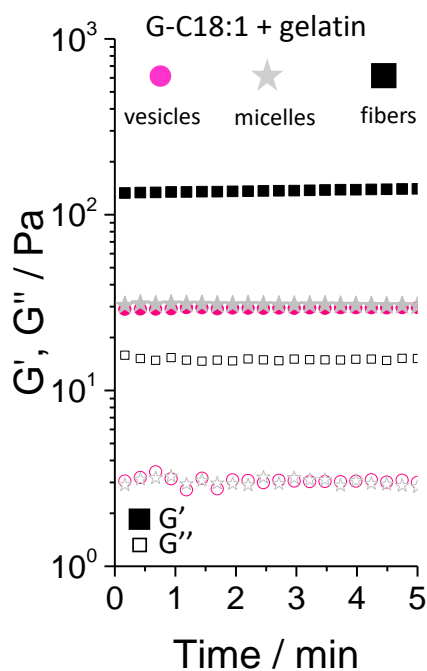

**Figure 2 -** Mechanical properties of hybrid gels probed by time-dependent oscillatory rheology (f= 1 Hz; γ= 0.1 %). Samples are prepared in mQ-grade water and composed of gelatin (2 wt%) mixed with either one of the three G-C18:1 (2 wt%) phases, micelles (pH= 8), vesicles (pH= 6) or fibers, (pH= 8, $[Ca^{2+}]$= 33.5 mM).

SAXS was employed to draw a clearer picture of the hybrid hydrogel's structure in comparison to the single components. Figure 3b presents the typical SAXS profile of the control and hybrid hydrogels. In the low-q regime (q< ~0.04 Å$^{-1}$), {F}G-C18:1 fibrillar hydrogels show a low-q slope with a -2 dependency of intensity (log-log scale) and a crystalline structure within the fiber network. The -2 dependency, typical for either mass fractals or flat structures,[39–41] characterizes {F}G-C18:1 gels and its origin has been discussed in detail elsewhere.[24] It can be explained by both the flat cross-section of the fibers but most likely by the raft-like, side-by-side, association of the fibers themselves. On the other hand, the scattering profile of gelatin follows a -1.7 dependency of Log(I)-Log(q), a value classically measured for swollen polymer chains,[39] over practically the entire q-range. It should be noted that the scattering profile of gelatin in water (Figure 3b) and gelatin prepared in a calcium solution (not shown) are comparable and suggest no major effects of calcium ions on the gelation of gelatin.

Interestingly, the scattering profile was -2.7 in the {F}G-C18:1/gelatin hybrid gel, a typical exponent observed for mass fractals,[39] and possibly suggesting inhomogeneities[42] at larger scale. These could be due to many factors like mixing, temperature fluctuations or, in this specific case, effects due to the mixing of two different networks. It could also be due to a simple arithmetic contribution of each signal. In the high-q regime, for q> ~0.1 Å$^{-1}$, the profile of {F}G-C18:1/gelatin appeared to correspond to the summation of the signals of the individual controls, as discussed below.

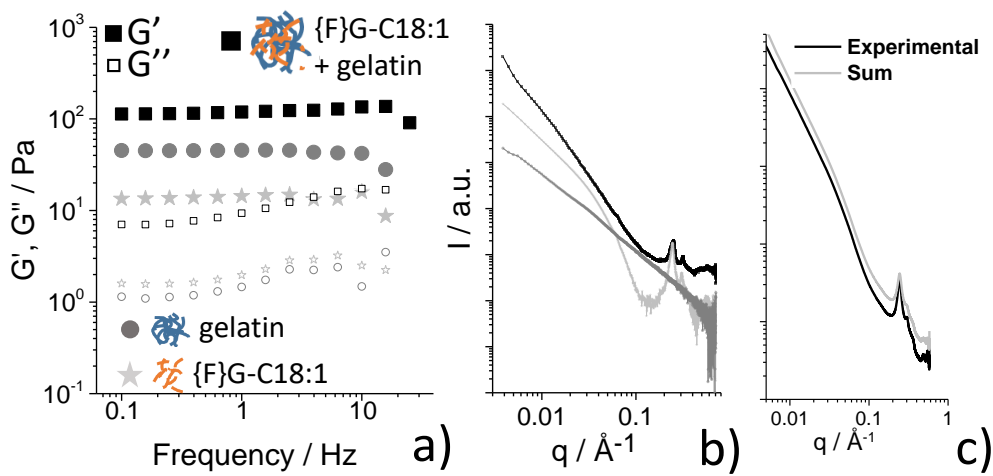

**Figure 3** – (a) Frequency-dependent (γ= 0.1 %) oscillatory rheology experiments and corresponding (b) SAXS profiles of {F}G-C18:1 (light grey), gelatin (grey), and hybrid {F}G-C18:1/gelatin (black) gels (C$_{gelatin}$= C$_{G-C18:1}$= 2 wt%, pH 8, [Ca$^{2+}$]= 33.5 mM). (c) Arithmetical sum (grey) of {F}G-C18:1 and gelatin SAXS profiles compared to the experimental profile of the hybrid {F}G-C18:1/gelatin hydrogel sample (black).

Figure 3c shows that the experimental SAXS signal of the hybrid gel is in very good agreement with the arithmetical sum of each individual's SAXS profile. Within the limits of SAXS, which is a statistical technique, this result can be taken as a proof of the interpenetration of SAFiN and polymer networks, and not of a mutual interaction. This result is also in agreement with our previous work, where the SAXS signal was profoundly modified when G-C18:1 was associated only with positively-charged polyelectrolytes, but not to gelatin in a broad pH range.[28] Figure 4 presents the rheo-SAXS experiment probing elastic behavior of the {F}G-C18:1/gelatin gel under strain sweep: about 80% of G' is recovered within 30 s, and at least over the three cycles of the experiment (Figure 4b). According to the corresponding scattering analysis (Figure 4a), which shows that SAXS profiles *(1)* (initial gel) through *(3)* (after two and three cycles) are superimposed, the structure of the hybrid gel is not affected.

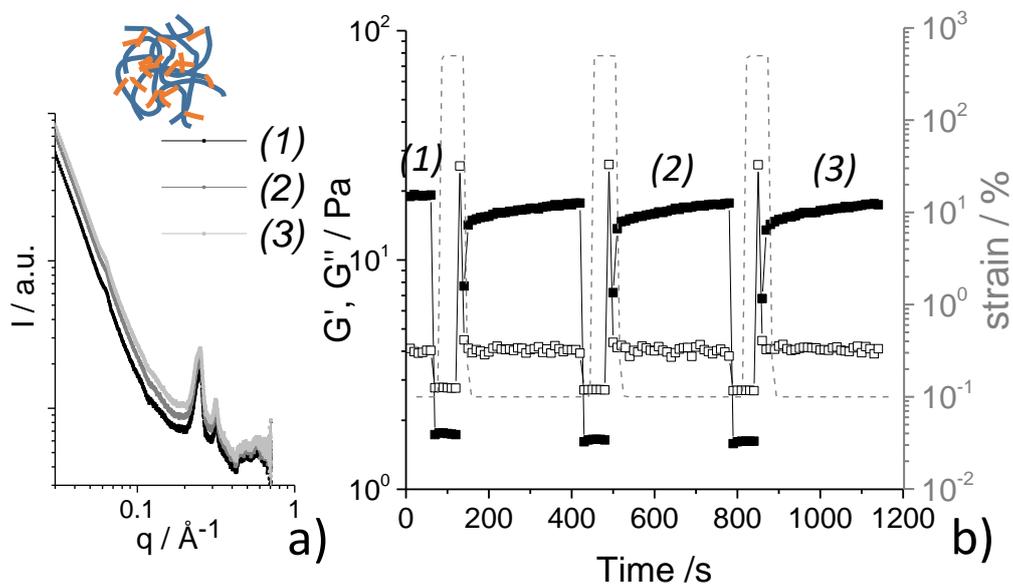

**Figure 4** – (a, b) Rheo-SAXS experiment (f= 1 Hz; γ= 0.1 %) probing the (b) elastic properties (oscillatory rheology) and (a) structures (SAXS) of {F}G-C18:1/gelatin hybrid gels shown in Figure 3 as a function of a strain cycle (0% - 100%).

*Hybrid {F}G-C18:1/alginate gels*

Alginate is a biopolymer whose hydrogelation process is triggered by the addition of $Ca^{2+}$ ions.[43] No gelation occurred when mixing a $Ca^{2+}$-free alginate solution with a G-C18:1 solution, either in a micellar or vesicle phase, , but gelation was immediate (Figure 5a) when adding a source of $Ca^{2+}$ to the alginate-micelles solution at basic pH, with G׳ above 100 Pa.

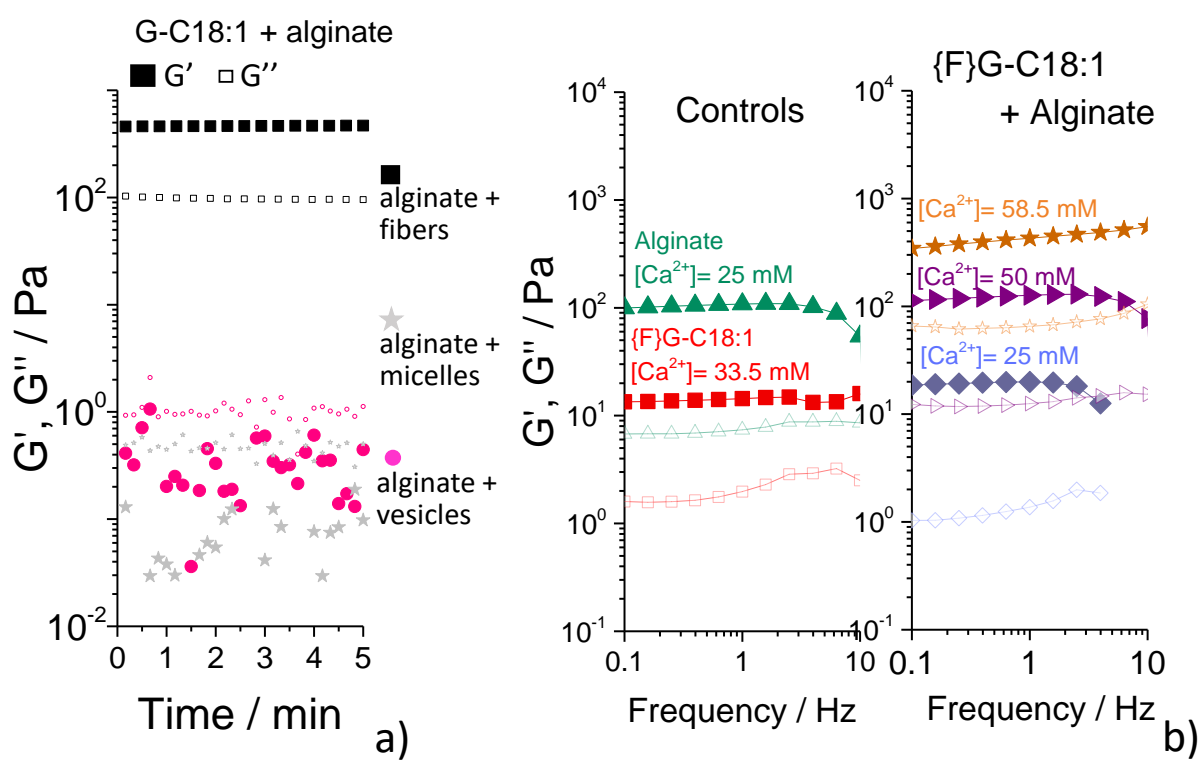

**Figure 5** – (a) Mechanical properties of hybrid gels probed by time-dependent oscillatory rheology (f= 1 Hz; γ= 0.1 %). Samples are prepared in mQ-grade water and composed of alginate (1 wt%) mixed with either one of the three G-C18:1 (2 wt%) phases, micelles (pH= 8), vesicles (pH= 6) or fibers, (pH= 8, [Ca2+]= 33.5 mM). (b) Frequency-dependent (γ= 0.1 %) oscillatory rheology experiments recorded on {F}G-C18:1/alginate hybrid gel ($C_{alginate}$= 1 wt%, $C_{G-C18:1}$= 2 wt%, pH 8) and corresponding controls (alginate and {F}G-C18:1) as a function of calcium concentration.

It is not surprising that alginate/micelle and alginate/vesicle hybrid samples are liquid solutions, alginate being known to form gels only in the presence of calcium.[43–46] Simultaneous responsivity to $Ca^{2+}$ ions for both alginate and G-C18:1 makes this sample particularly interesting. The G-C18:1/alginate system involves two partners, both containing carboxylic acid groups, whose gelation is triggered by calcium, thus creating competition. Figure 5b presents the mechanical properties of each component's solution in the presence of a calcium source: {F}G-C18:1 (2 wt%, 33.5 mM $CaCl_2$) and alginate (1 wt%, 25 mM $CaCl_2$). Upon addition of 25 mM $CaCl_2$ to the mixed alginate and G-C18:1 sample, the hybrid {F}G-C18:1/alginate hydrogel took on the mechanical properties of {F}G-C18:1 alone. When providing an additional source of calcium (50 mM), the hybrid gel took on the mechanical properties of the pure alginate gel. This simple experiment, which could be further verified by more advanced microcalorimetry tests, suggested that {F}G-C18:1 sequestrates calcium ions and possibly forms gels faster than alginate, which reacts with calcium only afterwards. Interestingly, the mechanical properties of the hybrid system could be improved by adding excess calcium, here increased up to 58.5 mM, and corresponding to the sum of the quantities required to gel each component.

The possibility of obtaining a stronger hybrid gel than each component taken separately demonstrated a cooperative effect between alginate and {F}G-C18:1, as also reported above for gelatin. Further experiments however showed that a threshold calcium concentration should not be exceeded to obtain a homogeneous gel. Under the present conditions, precipitation occurred at 100 mM $CaCl_2$.

The interaction between G-C18:1 and $Ca^{2+}$ was studied by calorimetry[24] and showed a bimodal interaction. An initial positive (endothermic) enthalpy variation ($\Delta H$= +6.01 kJ/mol) at low calcium content was attributed to non-specific interactions, presumably reflecting ion exchange (sodium-calcium) and a hydrophobic effect. A second, negative (exothermic), enthalpy variation ($\Delta H$= -11.17 kJ/mol) component was attributed to the complexation of calcium by G-C18:1. Interestingly, similar exothermic processes with enthalpy variation of the same magnitude ($\Delta H$= -15 kJ/mol and 11.6 kJ/mol, depending on the source of alginate) have been reported between alginate and $Ca^{2+}$.[43] Unfortunately, the similarity between the calorimetry data concerning the interaction between $Ca^{2+}$ and G-C18:1 or alginate did not help to determe which compound was preferred initially. Indeed, a possible calcium-mediated bridging interaction between G-C18:1 fibers and alginate cannot be excluded, thus possibly explaining the higher G´ values with a slight excess of $Ca^{2+}$ (58.5 mM, Figure 5b). Further experiments should be performed to verify this hypothesis.

Figure 6a-c presents the structural characterization of alginate glucolipid hydrogels studied by SAXS. The SAXS profile of alginate shows a -2.2 q-dependency of the intensity, a value generally associated with the scattering of an intermediate conformation between a globule (-3) and a Gaussian polymer chain (-2) rather than a free chain, as was found for gelatin.[39] The intensity in the {F}G-C18:1/alginate gel has a plain -2.0 dependency on q, thus reflecting the good interpenetration of the two networks.

Similar to what was found for the gelatin-based gels, the SAXS signature of {F}G-C18:1 is comparable to the signal of the hybrid {F}G-C18:1/alginate gel, in particular at q> 0.1 Å$^{-1}$ (Figure 6b). This reasonably suggests interpenetration between the SAFiN and polymer chains. Figure 6c,d presents the resulting rheo-SAXS experiment, probing the mechanical properties of the hybrid gel over several cycles of shear strain: 85% of G´ is recovered within 15 s at least over three consecutive cycles. As can be seen in Figure 6d, the structure of the gel is not affected, highlighted by the superimposable SAXS profiles corresponding to regions *(1)* through *(3)* of Figure 6c.

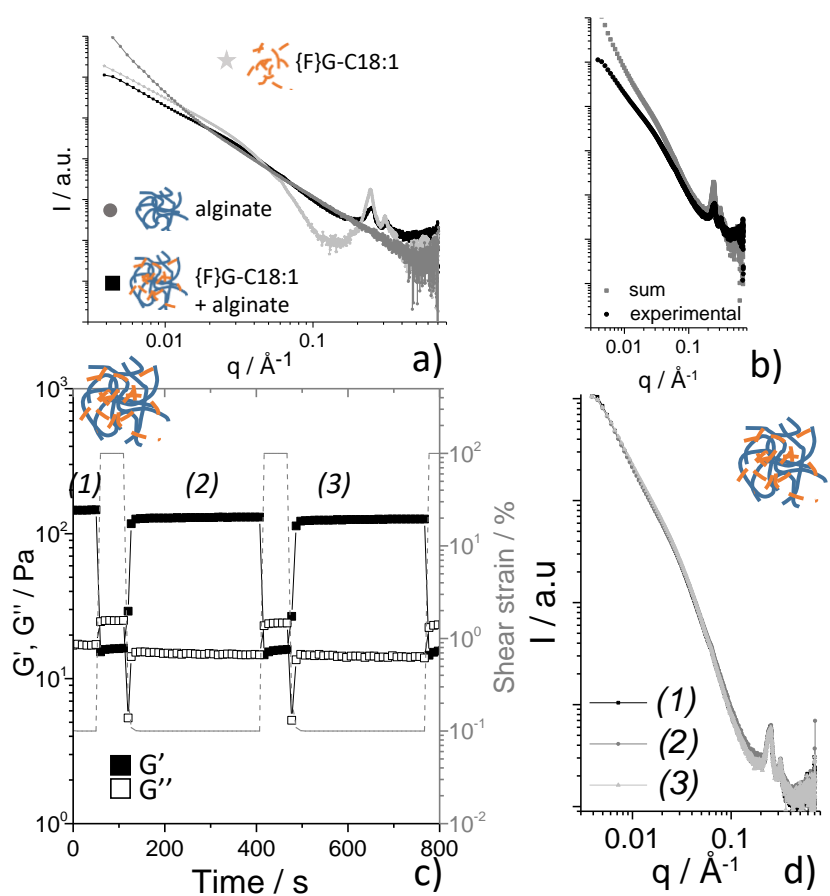

**Figure 6 - (a) SAXS profiles of {F}G-C18:1 (light grey), alginate (grey), and hybrid {F}G-C18:1/alginate (black) hydrogels ($C_{alginate}$= 1 wt%, $C_{G-C18:1}$= 2 wt%, pH 8, $[Ca^{2+}]$= 50 mM). (b) Arithmetical sum (grey) of {F}G-C18:1 and alginate SAXS profiles compared to the profile of the hybrid {F}G-C18:1/alginate hydrogel sample (black). (c, d) Rheo-SAXS experiment (f= 1 Hz; γ= 0.1 %) probing the (c) elastic properties (oscillatory rheology) and (d) structures (SAXS) of {F}G-C18:1/alginate hybrid gels as a function of a strain cycle (0% - 100%).**

*Hybrid {F}G-C18:1/chitosan gels*

Chitosan is a biopolymer, which is able to form a hydrogel above its $pK_a$. Each G-C18:1 phase, micellar, vesicular and fibrillar is combined with a chitosan gel, prepared by manually increasing of pH. Figure 7 shows that G'' > G' when chitosan is mixed with the vesicles, meaning that the latter do not provide any elastic properties to the system. In the case of a micellar solution, the viscous chitosan sample does not easily mix with the fluid micellar solution and the hybrid sample is too heterogeneous to measure reliable elastic properties.

However, when calcium is added to the latter micellar/chitosan solution, the sample immediately forms a homogeneous gel.

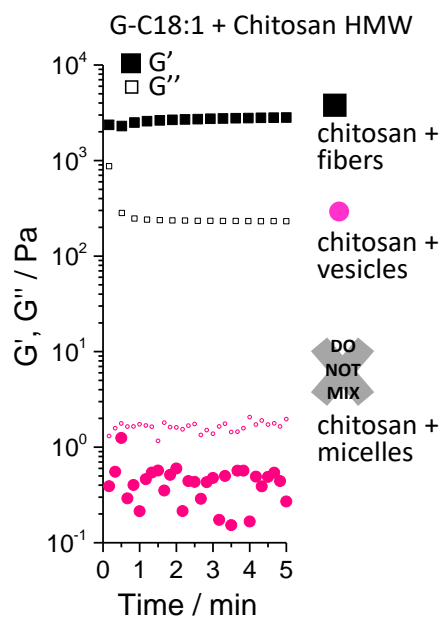

**Figure 7 - Mechanical properties of hybrid gels probed by time-dependent oscillatory rheology (f= 1 Hz; γ= 0.1 %). Samples are prepared in mQ-grade water and composed of chitosan (1 wt%) mixed with either one of the two G-C18:1 (2 wt%) phases, vesicles (pH= 6) or fibers, (pH= 8, [Ca$^{2+}$]= 33.5 mM). The micellar phase of G-C18:1 does not mix with the chitosan gel.**

Figure 8a presents the mechanical properties of the hybrid gel compared to those of each component alone: the combination of both chitosan and fibers provides a gel with enhanced mechanical properties, compared to the controls. Their corresponding SAXS profiles, given in Figure 8b, show the characteristic scattering pattern of {F}G-C18:1, indicating its presence and confirming its important role in the formation of the hybrid scaffold. The good matching between the SAXS patterns of {F}G-C18:1 and {F}G-C18:1/chitosan indicates the orthogonality between the two networks. Figure 8c presents the rheo-SAXS experiment associated with the step strain experiment performed on the hybrid gel: about 85% of the initial elastic properties are recovered within 30 s after releasing 100% strain over three cycles of the experiment. The superimposed SAXS data in Figure 8d, collected after each step-strain cycle, show that, similarly to alginate and gelatin-based hybrid gels, the structure is not affected (*(1)-(3)* profiles correspond to *(1)-(3)* on Figure 8c).

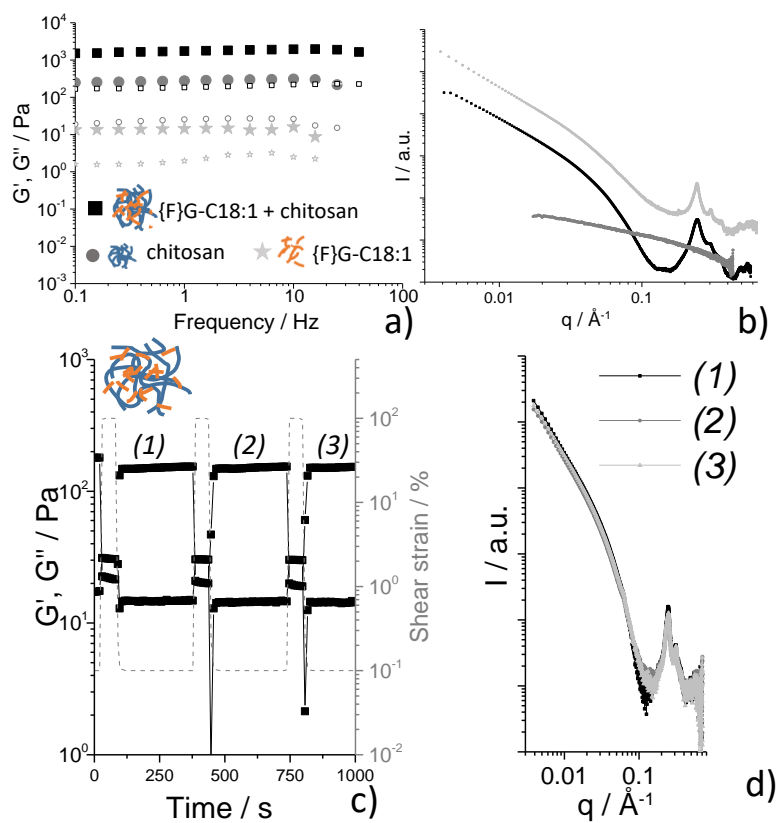

**Figure 8** – a) Frequency-dependent (γ= 0.1 %) oscillatory rheology experiments (G': full symbols; G'': empty symbols) and corresponding (b) SAXS profiles of {F}G-C18:1 (light grey), chitosan (grey), and hybrid {F}G-C18:1/chitosan (black) hydrogels ($C_{chitosan}$= 1 wt%, $C_{G-C18:1}$= 2 wt%, pH 8, $[Ca^{2+}]$= 33.5 mM). (c,d) Rheo-SAXS experiment (f= 1 Hz; γ= 0.1 %) probing the (c) elastic properties (oscillatory rheology) and (d) structures (SAXS) of {F}G-C18:1/chitosan hybrid gels as a function of a strain cycle (0% - 100%).

**Discussion**

Despite the differences in absolute values of G' for gelatin, alginate and chitosan hydrogels, the same tendency was observed upon the addition of a micellar, vesicular or fibrillar phase of G-C18:1. Micelles and vesicles both induced a decrease of the biopolymer hydrogel's mechanical properties, whereas fibers provided further strength to the biopolymer hydrogel. This is schematized in Figure 9. These effects are comparable with what is known about the effects of surfactants on the elastic properties of polymeric gels, but they have never been described for biosurfactant systems. In diluted surfactant/protein systems, changes in the protein's conformation, or even the denaturation of the protein, are classically observed in the vicinity of the CMC and attributed to hydrophobic effects between specific domains of the protein and the surfactant's aliphatic chains. Interestingly, milder effects, only above the CMC, have been reported when sophoro- and rhamnolipid biosurfactants replace classical anionic surfactants like sodium dodecyl surfactants.[47,48] Unfortunately, our experimental conditions are too far from these to be comparable: the CMC of the deionized (pH 10) form of G-C18:1 in its micellar phase was in the order of 0.2 wt% (Figure S 4), while the CMC of G-C18:1 at a more acidic pH, pH 6.5, when the vesicle phase formed, was in the order of 0.01 wt% (Figure S 4). These values are more than twenty times smaller than the GC18:1 concentration used in this work (2 wt%).

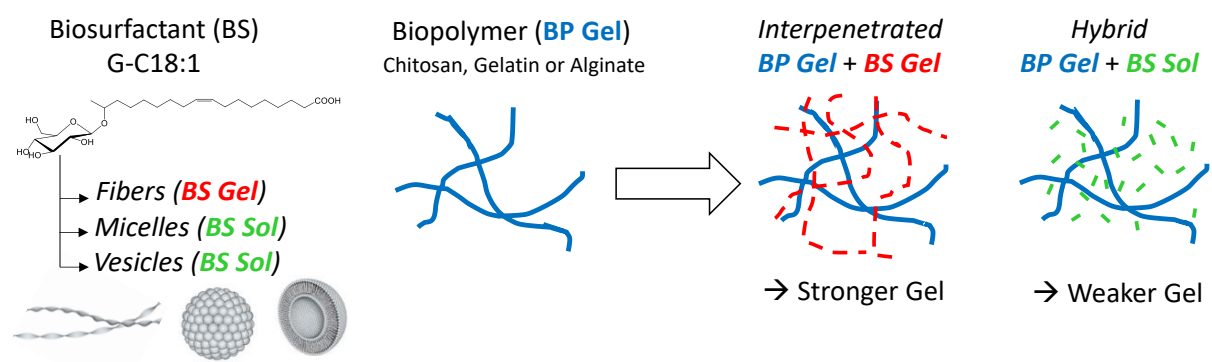

**Figure 9** – Scheme summarizing the behaviour of the hybrid G-C18:1/Biopolymer hydrogels according to the type of self-assembled G-C18:1 structure. Fibers strengthen the biopolymer hydrogel while micelles and vesicles weaken the gel.

When it comes to comparing the effect of amphiphile self-assembly on biopolymer gels, most of the data in the literature support the observations of this work: the presence of a second self-assembled fibrous network enhances the elastic properties of the biopolymer hydrogel,[12,49,50] as illustrated for instance by Nandi *et al.*, who synthetized folic acid/chitosan hydrogels with enhanced mechanical strength (from approximately 5 to 125 Pa)[49] or by Adams *et al.*, who controlled the viscosity, gelation time and thermal behavior of a non-gelling polymer dextran coupled to a pH-dependent low molecular weight dipeptide gelator.[12]

Concerning the effect of micellar solutions on the elasticity of biopolymer hydrogels, the literature agrees on the fact that the elastic properties are generally reduced.[51,52,53] As far as the effect of the vesicle phase is concerned, it seems to us that the literature lacks enough data to draw a general conclusion. Encapsulating vesicles in biopolymers is interesting to develop soft materials for biomedical applications, especially the encapsulation and release of various compounds.[54,55] However, the literature is mainly characterized by structural studies,[13] rather than the study and understanding of gel elasticity.[56,57]

From a mechanistic point of view, most authors formulate several hypotheses to explain the mechanism for reinforcing hybrid biopolymer/LMWG fiber hydrogels, but rarely support them with further experiments. Yang *et al.*, for example, assumed that interactions between the polymer and gel fibers or polymer-induced viscosity effects drive LMWG self-assembly: the presence of the polymer favors the bundling of the fibers, resulting in enhanced mechanical properties. Nandi *et al.* [49] assessed a key role of hydrogen bond interactions between chitosan and folic acid, at the origin of increased branching. According to them, the additive is responsible for lash-shaped branched networks in the LMWG gel, as detected from the TEM image of the chitosan/LMWG gel. The fibers' diameter in the hybrid gel is also significantly reduced with respect to LMWG gel. Both factors suggest the increase in the surface force of the fibrillar network, entrapping the solvent molecules more tightly. In these systems involving micelles or vesicles, the literature provides arguments when discussing mesh size in relation to the elastic properties of the gel.[58–60] Wang *et al.* for instance reported an increase in the mesh size of hydroxyethylcellulose (HEC) and hydrophobically modified hydroxyethylcellulose (HMHEC) hydrogels with increasing SDS micelle concentrations, while the correlation length was found to be decreased in the meantime.[11]

The SAXS profiles corresponding to the hybrid {F}G-C18:1/biopolymer hydrogels show that the high *q* region, which probes the scale between 1 and 10 nm, match the arithmetical sum of the signals corresponding to the individual components, the biopolymer and {F}G-C18:1. This strongly suggests that neither the polymer network nor the fiber's structure are perturbed at length scales below 10 nm. This fact excludes variations in the morphology, cross-section and structure of the {F}G-C18:1 fibers. For the larger scales at lower *q* values, SAXS profiles can be fitted with the Ornstein-Zernike (OZ) equation, generally employed for swollen gels[42] and giving access to the mesh size of the biopolymer gel network. According to previous data, the presence of micelles is assumed to have an impact on the mesh size and, consequently, the mechanical properties of polymeric hydrogels.[58–60] In contrast, the presence of fibers enhances the mechanical properties in the hybrid gel, although the impact on the polymer's mesh size is less clear.

The mesh size, $\xi$, can generally be obtained from the OZ equation ($I(q) = \frac{I(0)}{(1+q^2\xi^2)}$) if I(0), the scattering intensity at q= 0, is known. Since this is not the case for any of the SAXS profiles recorded in this work, only a qualitative approach should be employed. According to the analysis described in Figure S 5 and Table S 1 in the Supporting Information, the biopolymer network does not seem to be particularly modified by the presence of the second interpenetrated self-assembled fibrous network. Other mechanisms, such as that mentioned by Qin *et al.*,[61] suggesting that fiber-reinforced alginate hydrogels are able to absorb water from a second interpenetrated network, could probably help understanding, although further experiments are required. When it comes to micelle-containing hydrogels, the analysis in Table S 1 shows several differences between the gelatin control gel and the hybrid G-C18:1/gelatin, thus indicating that the micelles could have an impact on the gelatin network, as reported for other polymer-surfactant hydrogels.[11,62]

Although it could be argued that SAXS may not be appropriate to understand the evolution of the elastic properties of hybrid hydrogels, it would not be unreasonable to study this and other similar systems under different experimental conditions. The use of ultrasmall-angle X-ray and neutron scattering (USAXS, USANS) could be helpful to verify that a plateau in the I(q) is eventually reached.[42] Contrast-matching SANS could also be employed to isolate the contributions of the polymer and amphiphile to the hybrid network, as done elsewhere.[63] Finally, the use of model-dependent fitting procedures, although cumbersome for such complex hybrid systems, could also help identifying the contribution of these key parameters involved in the architectural changes of the network.

**Conclusion**

Fully biobased hydrogels with tunable mechanical properties were obtained by mixing a glycolipid biosurfactant and a biopolymer (gelatin, alginate or chitosan). The glycolipid G-C18:1 had a pH-dependent and ion-dependent phase behavior at room temperature under diluted conditions. At basic pH in the presence of calcium, G-C18:1 formed self-assembled fibers which improve the mechanical properties of the biopolymer gel by about one order of magnitude. The addition of vesicles (pH< 7) or micelles (pH> 7) to the biopolymer hydrogel resulted in a gel with slightly lower mechanical strength. These observations seem to be generalized to all the biopolymers tested here and could be reproduced under non-equilibrium conditions by means of pH-resolved *in situ* SAXS experiments.[64]

Both the increase and decrease of mechanical properties were tentatively explained through the possible evolution of the biopolymer's mesh size. Despite the fact that the SAXS data did not reach a plateau at low q-values, which was necessary to quantify the mesh size, the Ornstein-Zernike plots suggested that the insertion of micelles contributed to increasing the mesh size of the biopolymer hydrogel, while the fibers interpenetrated the biopolymer gel without specific interactions.

**Acknowledgements**

We thank the Soleil Synchrotron facility for making available the Swing beamline and financial support (proposal No. 20200532). We also thank Ghazi Ben Messaoud (DWI-Leibniz Institute for Interactive Materials, Aachen, Germany) for helpful discussions. We are grateful to Dr. S. Roelants and Prof. W. Soetaert at Gent University for shipping the glycolipid. Sorbonne Université (contract No. 3083/2018) is acknowledged for its financial support of CS. The authors gratefully acknowledge the French ANR, Project No. SELFAMPHI - 19-CE43-0012-01.

**Supporting Information content**

Figure S 1 shows the time dependence of storage and loss moduli of the G-C18:1 fiber hydrogel and {F}G-C18:1 (2 wt%) at pH 8. Figure S 2 contains the time dependence of storage and loss moduli of gelatin (2 wt%), alginate (1 wt%, 25 mM $CaCl_2$) and chitosan HMW (1 wt%). Figure S 3 displays a repeatability experiment of the storage modulus for several freshly-prepared samples of the hybrid {F}G-C18:1/gelatin hydrogels. Figure S 4 shows the CMC of G-C18:1 in water at pH 10 and 6.5. Figure S 5 shows the Ornstein-Zernike plots for G-C18:1/biopolymer hydrogels. Table S 1 shows the calculations for the mesh size.

**TOC Graphics**

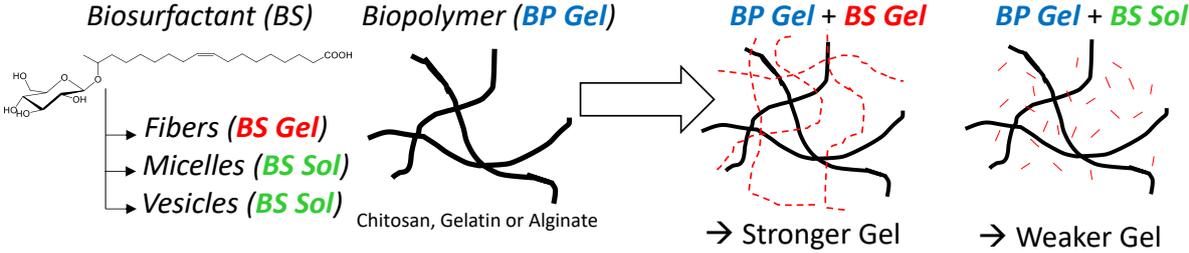

# Interpenetrated biosurfactant-biopolymer orthogonal hydrogels: the biosurfactant's phase controls the hydrogel's mechanics

Chloé Seyrig,[a] Alexandre Poirier,[a] Javier Perez,[b] Thomas Bizien,[b] Niki Baccile[a,*]

[a] Sorbonne Université, Centre National de la Recherche Scientifique, Laboratoire de Chimie de la Matière Condensée de Paris , LCMCP, F-75005 Paris, France
[b] Synchrotron SOLEIL, L'Orme des Merisiers Saint-Aubin, BP 48 91192 Gif-sur-Yvette Cedex

\* Corresponding author:
Dr. Niki Baccile
E-mail address: niki.baccile@sorbonne-universite.fr
Phone: +33 1 44 27 56 77

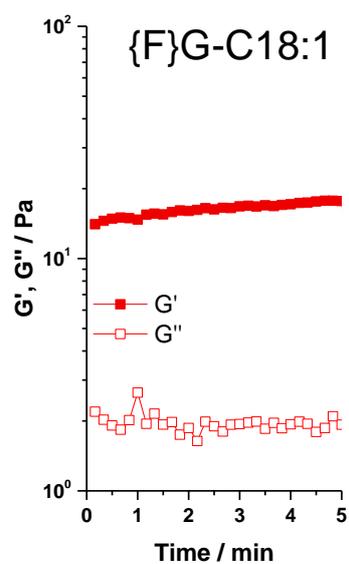

**Figure S 1 – Time dependence of storage and loss moduli (f= 1 Hz, γ= 0.1 %) of the G-C18:1 fiber hydrogel, {F}G-C18:1 (2 wt%), at pH 8 containing [Ca$^{2+}$]= 33.5 mM. Moduli are followed after loading the gel on the rheometer's geometry. Analysis is performed under oscillatory conditions in the linear viscoelastic regime.**

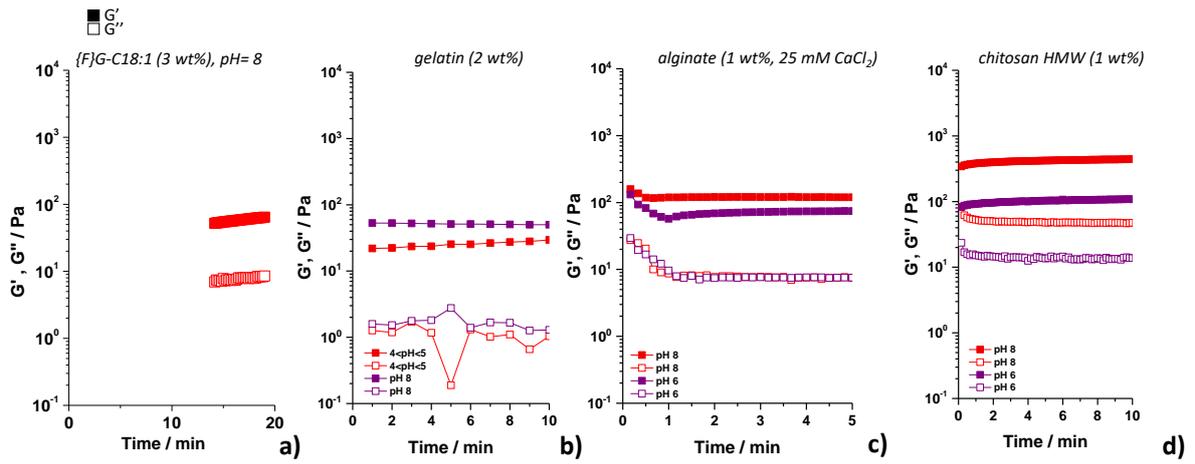

**Figure S 2** – Time dependence of storage and loss moduli (oscillatory rheology, LVER, f= 1 Hz, γ= 0.1 %) of a) gelatin (2 wt%), b) alginate (1 wt%, 25 mM CaCl$_2$) and c) chitosan HMW (1 wt%) in water at acidic and basic pH. Moduli are followed after loading the gel on the rheometer's geometry.

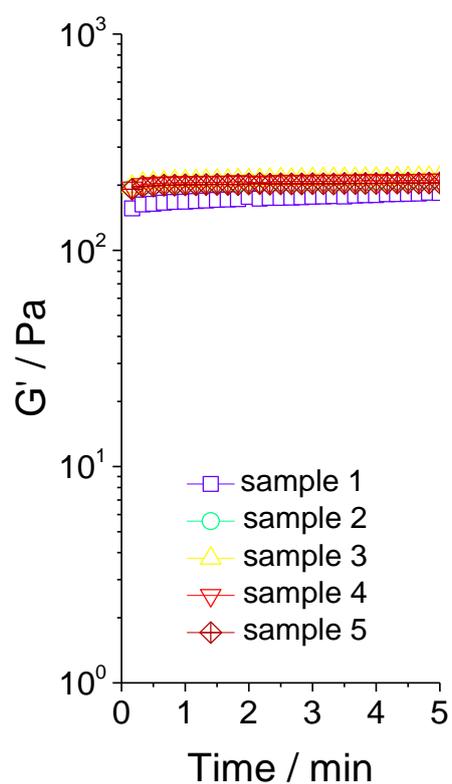

**Figure S 3 – Repeatability experiment. Storage modulus recorded over time (LVER, f= 1 Hz, γ= 0.1 %) for several freshly-prepared samples of the hybrid {F}G-C18:1/gelatin hydrogels ($C_{gelatin}$= $C_{G\text{-}C18:1}$= 2 wt%, pH= 8, $[Ca^{2+}]$= 33.5 mM).**

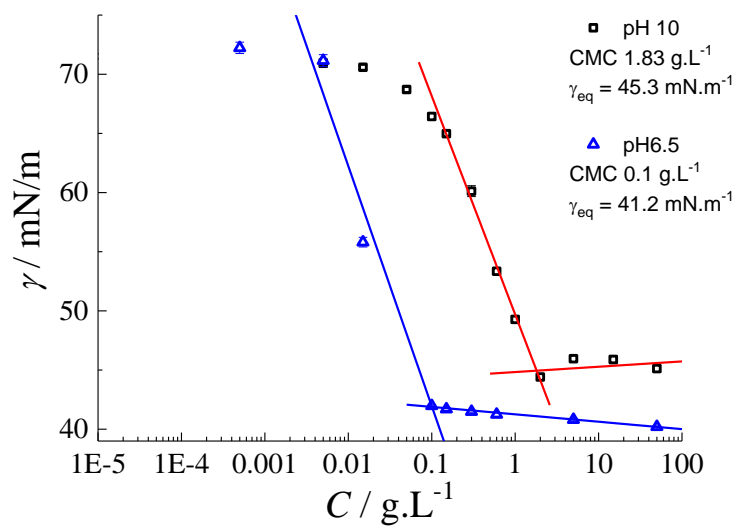

**Figure S 4 – Evolution of the static surface tension with G-C18:1 concentration measured at room temperature (23°C). At pH 10, G-C18:1 is fully ionized and in a micellar phase while at pH 6.5, G-C18:1 is partially ionized and in a vesicle phase.**

**Ornstein-Zernike analysis**

The Ornstein-Zernike equation is (Eq. S1)

$$I(q) = \frac{I(0)}{(1+q^2\xi^2)} \qquad \text{Eq. S1}$$

Which, rearranged, gives

$$\frac{1}{I(q)} = \frac{1}{I(0)} + q^2\frac{\xi^2}{I(0)} \qquad \text{Eq. S2}$$

with ξ being the mesh size of the polymer gel and I(0) the scattering intensity at q= 0. Plotting Eq. S2 as 1/I(q) against $q^2$, one could estimate ξ from the slope ($\xi^2$/I(0)), if I(0) is known. However, this is unfortunately seldom the case, as the scattering plateau at q= 0 is generally not achieved for many gel and colloidal systems. As it can be easily observed, none of the SAXS profiles reported in this work reaches a plateau at q= 0, making it impossible to unambiguously quantify ξ. Despite this major drawback, one could still compare the values of the slopes in a relative fashion, supposing that I(0) is comparable across samples of similar structure. Figure S 5 and Table S 1 provide the Ornstein-Zernike plots and corresponding $\frac{\xi^2}{I(0)}$ values for the hybrid gelatin and alginate hydrogels, compared to the controls. The values of $\frac{\xi^2}{I(0)}$ for the fiber-containing hydrogels, may them contain gelatin or alginate, are comparable to the respective controls within the incertitude of the fit. When it comes to micelle-containing hydrogels, the gelatin structure shows a smaller $\frac{\xi^2}{I(0)}$ if compared to the control. However, lacking the values of I(0), it is impossible to establish a quantitative correlation between $\frac{\xi^2}{I(0)}$, the mesh size and the elastic properties of the hydrogel.

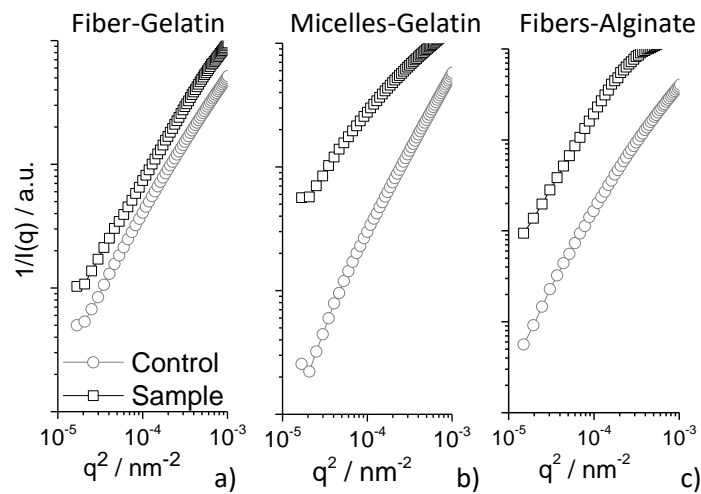

**Figure S 5** - Ornstein-Zernike plots of the profiles corresponding to hybrid G-C18:1/biopolymer hydrogels and free biopolymer controls. Controls (black squares) are a) $Ca^{2+}$ + gelatin, b) gelatin and c) $Ca^{2+}$ + alginate. Samples are intended as the biopolymer network inside the hybrid G-C18:1 – biopolymer medium. The signal of the sample is obtained by subtracting the signal of the G-C18:1 phase to the signal of the hybrid G-C18:1 – biopolymer system: a) [{F}G-C18:1 + gelatin] - {F}G-C18:1; b) [{M}G-C18:1 + gelatin] - {M}G-C18:1; c) [{F}G-C18:1 + alginate] - {F}G-C18:1.

**Table S 1** – List of the $\frac{\xi^2}{I(0)}$ defined in Eq. S1 and Eq. S2 and obtained as the slope of the Ornstein-Zernike plots in Figure S 5

| G-C18:1 phase | Sample | $\frac{\xi^2}{I(0)}$ |
|---|---|---|
| Fiber | Control ($Ca^{2+}$ + gelatin) | 1.19 |
| Fiber | Gelatin in {F}G-C18:1/gelatin hybrid | 1.19 |
| Micelle | Control (gelatin) | 1.33 |
| Micelle | Gelatin in {M}G-C18:1/gelatin hybrid | 0.80 |
| Fibers | Control ($Ca^{2+}$ + alginate) | 1.65 |
| Fibers | Alginate in {F}G-C18:1/alginate hybrid | 1.61 |